**Raitina M.Y., Pustovarova A.O., Pokrovskaya E.M.**

# The educational process organization in the distance learning model: problems and features

This article discusses the current issues of providing the educational process in a remote format on the example of the Tomsk State University of Control Systems and Radioelectronics (TUSUR). On the basis of the conducted research, the problems and features of distance learning are identified and analyzed. It is concluded that it is necessary to modernize education on the basis of constant monitoring of the effectiveness of providing distance learning, taking into account the opinions of all participants in the educational process.

The sudden transition of universities to a distance learning format within the framework of anti-epidemic measures revealed a number of problems and features of distance learning (DL), which have a significant impact on its quality. Thus, such educational formats as e-learning, distance forms of interaction, and a hybrid educational model have become significant. Therefore, today the identification and analysis of the problems and features of the above-mentioned forms of education in the educational process of the university are becoming particularly relevant [1, 2].

To identify these factors, a study was conducted among TUSUR students in June 2020. It was attended by 240 students of various faculties of 1-4 courses. As a result of the monitoring, the following results were obtained.

When asked about changes in the quality of their education, the respondents ' opinions were distributed as follows

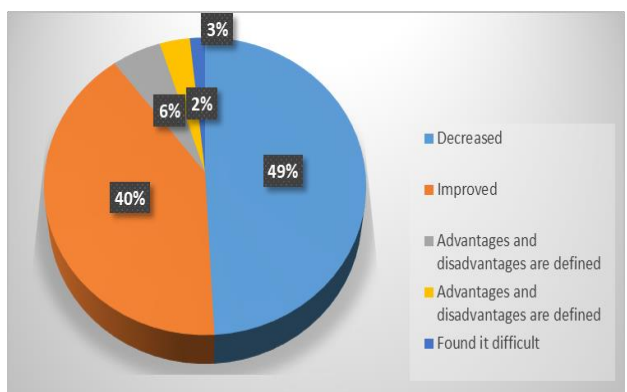

Fig. 1. Quality of education

The main reasons for the deterioration of the quality of education, while choosing this option, are:

1) the absence of direct contact with the teacher – 30 persons;

2) overall DL complexity, inability to understand yourself in the material, the complexity of learning materials in electron form – 32 persons;

3) poor training, the deterioration of the quality of the lecture material, or even the lack of lectures as such that leads to self-obtaining information from the Internet and difficulties with understanding the material – 20 persons;

4) difficulties with self-discipline, savorani-organization and General maladjustment – 7 persons;

5) lack of preparation for the transition to pre-school education for both students and teachers – 6 persons;

6) technical problems – 4 persons.

19 persons noted the deterioration in the quality of training, without specifying its cause.

According to the respondents, the main motive for improving the quality of education was comfort and convenience of learning.

The change in the academic load due to the transition to the distance learning format was assessed by the respondents as follows:

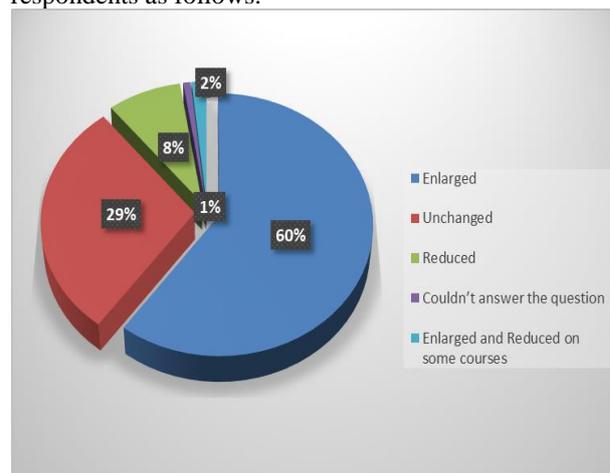

Fig. 2. Changing the volume of training work

Respondents, who noted an increase of the volume of training work, identified the following reasons for this increase:

1) the development of educational material (lack of understanding, long waiting for the answer / help from the teacher, etc.) – 63 persons;

2) the teachers began to ask more training material – 34 persons;

3) uneven distribution of the volume of training work, leading to trouble, due to delays in preparation of assignments teachers - 6 persons.

4) fatigue from constant work at the PC, not being able to change the situation, - 5 persons;

36 persons did not specify the reason for the increase.

Respondents who indicated that their volume of training work decreased explained this decrease by the following reasons:

1) the absence of lectures – 6 persons;
2) the absence of need to go to classes – 4 persons;
3) independent time allocation – 3 persons;

7 persons did not specify the reason for the decrease.

The request to evaluate the quality of the training materials provided and the educational environment caused the following responses:

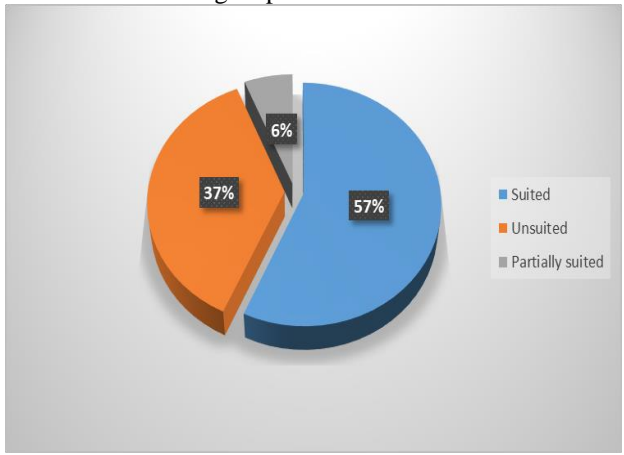

Fig. 3. Evaluate the quality

Among the things that do not suit the students, were named:

1) absence/lack of lectures and other online classes – 15 persons;
2) lack of live communication with teachers or at least operational communication with them – 18 persons;
4) hastily, poorly, without explanations, materials- 23 persons;
5) technical shortcomings of Moodle and other platforms-13 persons;
6) the DL organization does not suit - 7 persons;
7) the ratio of the cost of training and the quality of education in connection with the quarantine – 3 persons.

10 persons didn't satisfied, but the reason is not specified.

On the question of the ability of distance education to develop creativity and instill useful skills, the respondents expressed the following opinion:

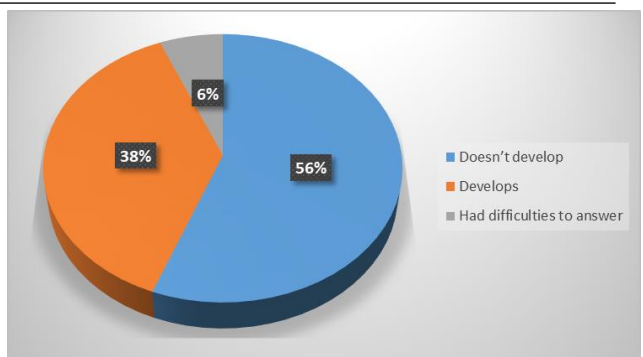

Fig. 4. Development of creative abilities

Respondents who responded positively to this question noted the development of distance learning in their countries.:

1) self-organization, self-discipline – 29 persons;
2) creativity, self-education – 17 persons;
3) time management, rational distribution of time – 11 persons;
4) skills of working with information – 15 persons;
5) skill of using various programs, Internet services 13 persons;
6) communication skills with teachers, classmates – 5 persons.

1 person did not specify what exactly distance education develops for him.

Among the various types of independent work used in the educational process in the distance format, respondents were asked to indicate the most effective, in their opinion. Below are the answers:

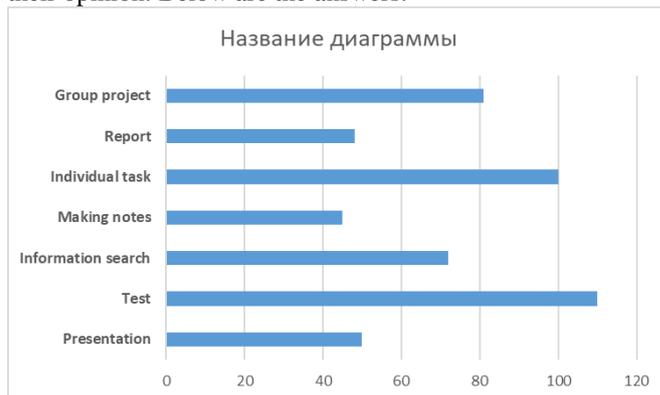

Fig. 5. The most effective types of independent work

There are students ' answers about which format of training – full-time or distance is more convenient to them:

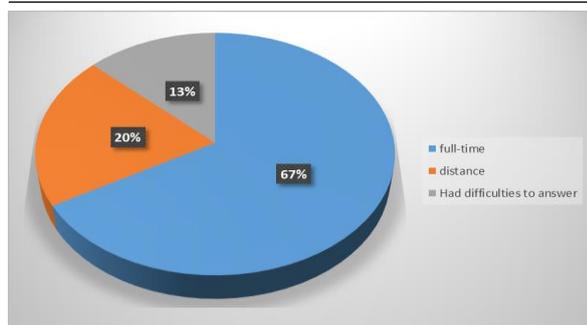

Fig. 6. Preferred training format

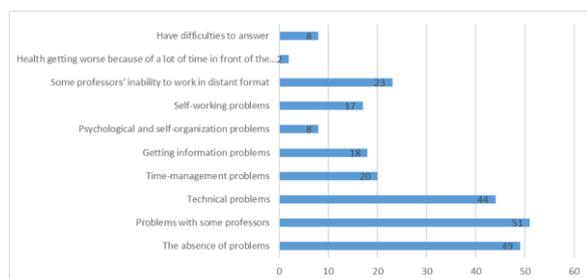

Fig. 7. Difficulties faced by students in distance learning

Due to the previously stated problems of regular communication with a number of teachers, the respondents were also asked to answer the question of whether their teachers are available for prompt answers. The responses were as follows:
- Yes the major part (40%);
- Some people only (39%);
- A few people (17%);
- No not accessed (4 %).

The final question of the survey was the question about the respondents ' proposals for organizing ditance learning. The respondents made various suggestions, ranging from non-constructive ones (for example, the complete abolition of distance learning) to technical comments on improving certain elements of the Moodle platform. Below are the main wishes of the respondents to the applicants in the context of the ongoing pandemic:

1) Conduct classes in the format of videoconference and webinars to present lectures in different forms and video, and text – 11 persons

2) to be in touch with students to answer questions and work quickly, to articulate the conditions and deadlines, disassemble admitted students mistakes – 10 persons;

3) to develop training manuals for use in a remote format, i.e., with detailed instructions and rashes-up of the implementation of obscure places, to complement the e-learning content inaccessible literature – 6 persons;

4) Rationally and evenly distribute training tasks throughout the semester in order to avoid overloads at the end of the semester and provide enough time for their implementation – 4 persons.

Thus, to ensure the quality of education and maintain its main purpose the necessary regulation of lifelong learning, taking into account external challenges (for example, conditions of the pandemic) in a hybrid educational model that actualizes constant monitoring of the assessment of the level of implementation and effectiveness of distance learning taking into account the views of all interested persons – participants of the educational process [3-5].

*Acknowledgements*

The research was supported by the basic part of the state assignment «Science» (scientific project code FEWM-2020-0036).

———————————————————————

**Margarita Y. Raitina**
PhD in Philosophy, associate professor of the philosophy and sociology department
Tomsk State University of Control Systems and Radioelectronics (TUSUR)
40, Lenina st., Tomsk, Russia, 634045
ORCID: 0000-0002-2381-3202
Phone: +7 (382-2) 70-15-90
Email: raitina@mail.ru

**Anna O. Pustovarova**
Senior Lecturer of the philosophy and sociology department
Tomsk State University of Control Systems and Radioelectronics (TUSUR)
40, Lenina st., Tomsk, Russia, 634045
Phone: +7 (382-2) 70-15-90
Email: anna.o.pustovarova@tusur.ru

**Elena M. Pokrovskaya**
PhD in Philosophy, head of the foreign languages department
Tomsk State University of Control Systems and Radioelectronics (TUSUR)
40, Lenina st., Tomsk, Russia, 634045
ORCID: 0000-0002-2381-3202
Phone: +7 (382-2) 70-15-21
Email: pemod@yandex.ru